\documentclass[]{article}
\usepackage[fleqn]{amsmath}  

\title{Changes to the $t$-$J$ Model From Inclusion of Two-Site Nearest-Neighbor Coulomb Interactions}
\author{Tao Sun\\
\small{tao@taosphysics.net}}

\begin{document}

\maketitle

\begin{abstract}

To date, the Hubbard model and its strong coupling limit, the $t$-$J$ model, serve as the canonical model for strongly correlated electron systems in solids.
Approximating the Coulomb interaction by only the on-site term (Hubbard $U$-term), however, may not be sufficient to describe the essential physics of interacting electron systems.
We develop a more complete model in which all the next leading order terms besides the on-site term are retained.
Moreover, we discuss how the inclusion of these neglected interaction terms in the Hubbard model changes the $t$-$J$ model.

\end{abstract}

\section{Introduction}

It is generally believed \cite{anderson} that some essential physics of strongly correlated electron systems in solids, such as high-temperature superconductors \cite{bednorz}, can be described by the Hubbard model \cite{hubbard,gutzwiller}
\begin{equation}\label{eq:hubbard}
\begin{aligned}{}
H=-t\sum_{ij,\sigma}\gamma_{ij}c^{\dagger}_{i \sigma}c_{j \sigma}
+ U \sum_{i} n_{i\uparrow}n_{i\downarrow},
\end{aligned}
\end{equation}
where $c^{\dagger}_{i \sigma}$ and $c_{i \sigma}$ are the creation and annihilation operators for an electron of spin $\sigma$ at site $i$, and $n_{i\sigma}=c^{\dagger}_{i \sigma} c_{i\sigma}$, the corresponding number operator of electrons.
In Eq (\ref{eq:hubbard}),
$\gamma_{ij}=1$ for nearest-neighbor (n.n.) sites $i,j$, and $0$ otherwise, which restricts the summation over the nearest-neighbor pairs.
The $t$-term describes the kinetic energy due to hopping between adjacent sites, and the $U$-term is the on-site approximation of the Coulomb interaction between electrons.
Indeed, the Hubbard model offers the simplest way to describe both the band motion and atomic features of the interacting electron systems in solids.

In the Hubbard model, there are four possible states on the lattice sites: empty state $|0\rangle$, singly occupied states $|\uparrow\rangle$ (spin up) and $|\downarrow\rangle$ (spin down), and doubly occupied state $|\uparrow \downarrow\rangle$. 
Apparently, the double occupancy state $|\uparrow \downarrow\rangle$ is not favorable when $U$ is large. 
It is commonly believed that a model in which all the doubly occupied states are excluded can describe the low-energy physics of the system. 
Technically, such a singly occupied state system (with only empty and singly occupied sites) can be obtained by the Gutzwiller projection operator approach, which projects the Hubbard model to a subspace of the original Hilbert space \cite{fock}.
For large $U$, to the order of $J=2t^2/U$ in the perturbation calculation, this projecting scheme leads to the well-known $t$-$J$ model \cite{spalek,zhang}
\begin{equation}\label{eq:tj}
\begin{aligned}{}
H=-t\sum_{ij,\sigma}\gamma_{ij}\tilde{c}^{\dagger}_{i \sigma}\tilde{c}_{j \sigma}
+ J \sum_{ij} \gamma_{ij} (S_{i}\cdot S_{j}
-\frac{1}{4}\tilde{n}_{i}\tilde{n}_{j}),
\end{aligned}
\end{equation}
where 
$\tilde{c}^{\dagger}_{i \sigma}=
c^{\dagger}_{i \sigma}(1-n_{i\bar{\sigma}})$, 
$\tilde{c}_{i \sigma}=
c_{i \sigma}(1-n_{i\bar{\sigma}})$,
$\tilde{n}_{i}=\sum_{\sigma}
\tilde{c}^{\dagger}_{i \sigma}\tilde{c}_{i \sigma}$,
and
$S_i$ is the spin operator at site $i$.
Note that the projection operator also generates a so-called three-site term
that has been neglected in the $t$-$J$ model.
That is, Eq (\ref{eq:tj}) is in fact a model with only two-site terms.
The $t$-$J$ model can be viewed as a large $U$ limit of the Hubbard model, and as a prototype to describe the low-energy physics of the interacting electron systems. 
Using the standard Green function technique, one can indeed show that the $t$-$J$ model does exhibit a state with nonzero 
$\langle c_{i\downarrow}c_{j\uparrow}\rangle$, 
signaling the occurrence of superconductivity. 
Nonetheless, it is still somewhat difficult to obtain a more complete understanding of high-temperature superconductors from the $t$-$J$ model. 

The Hubbard model is the simplest approximation of the general Hamiltonian of interacting electron systems, in which all the interaction terms except the on-site term ($U$-term) are neglected.  
It is possible that the neglected interaction terms may play important roles in the understanding of the physics of the strongly correlated electron systems. 
Some important physics may be omitted in the Hubbard model.
As a result, the $t$-$J$ model given in Eq (\ref{eq:tj}), which is a projection to the subspace of singly occupied states of the Hubbard model, may not be sufficient to describe the essential physics of the interacting electrons.

In this paper, we will first describe the details of establishing a generalized Hubbard model, in which all of the two-site interaction terms between nearest-neighbor sites are retained.
We will then apply the Gutzwiller projection scheme to this more general model to obtain an effective Hamiltonian in the subspace of singly occupied states \cite{spalek0}.
Moreover, we will discuss how the inclusion of these neglected interaction terms in the Hubbard model modifies the $t$-$J$ model.

\section{The Model}

We now present a systematic derivation of a more complete effective Hamiltonian with all two-site interaction terms between nearest-neighbor sites in the subspace of singly occupied states.

\subsection{Bloch Representation}

The general Hamiltonian describing the dynamics of the electrons in solids can be expressed as follows \cite{hubbard,mahan,lzz}  in Bloch representation
\begin{equation}\label{eq:bloch}
\begin{aligned}{}
H &=\sum_{\textbf{k},\sigma}\epsilon_{\textbf{k}}
c^{\dagger}_{\textbf{k}\sigma}c_{\textbf{k}\sigma}
+\frac{1}{2}\sum_{\textbf{k}_{1}\textbf{k}_{2}
\textbf{k}^{'}_{1}\textbf{k}^{'}_{2},\sigma\sigma^{'}}
\langle\textbf{k}_{1}\textbf{k}_{2}|\frac{1}{r}|
\textbf{k}^{'}_{1}\textbf{k}^{'}_{2}\rangle
c^{\dagger}_{\textbf{k}_{1}\sigma}
c^{\dagger}_{\textbf{k}_{2}\sigma^{'}}
c_{\textbf{k}^{'}_{2}\sigma^{'}}
c_{\textbf{k}^{'}_{1}\sigma},
\end{aligned}
\end{equation}
where
$c^{\dagger}_{\textbf{k},\sigma}$ and $c_{\textbf{k},\sigma}$ are the creation and annihilation operators for electrons in the Bloch state with wave vector $\textbf{k}$ and spin $\sigma$, $\epsilon_{\textbf{k}}$ is the band energy, $\textbf{k}$ runs over the first Brillouin zone, and the matrix element
\begin{equation}
\begin{aligned}{}
\langle\textbf{k}_{1}\textbf{k}_{2}|\frac{1}{r}|
\textbf{k}^{'}_{1}\textbf{k}^{'}_{2}\rangle=e^2
\int d \textbf{x} d \textbf{x}^{'} 
\frac{\psi^{*}_{\textbf{k}_{1}}(\textbf{x})
\psi_{\textbf{k}^{'}_{1}}(\textbf{x})
\psi^{*}_{\textbf{k}_{2}}(\textbf{x}^{'})
\psi_{\textbf{k}^{'}_{2}}(\textbf{x}^{'})}
{|\textbf{x} - \textbf{x}^{'}|},
\end{aligned}
\end{equation}
where $\psi_{\textbf{k}}(\textbf{x})$ and 
$\psi^{*}_{\textbf{k}}(\textbf{x})$ 
are the Bloch functions of the energy band. 
In Eq (\ref{eq:bloch}), the first term represents the kinetic energies of the electrons in the energy bands, and the second term is the Coulomb interaction energy of electrons. 

\subsection{Wannier Representation}

The Hamiltonian in Eq (\ref{eq:bloch}) can be transformed to its Wannier representation form
\begin{equation}\label{eq:wannier}
\begin{aligned}{}
H &=\sum_{ij,\sigma}T_{ij}c^{\dagger}_{i \sigma}c_{j \sigma}
+\frac{1}{2}\sum_{ijkl,\sigma\sigma^{'}}
\langle ij|\frac{1}{r}|kl\rangle
c^{\dagger}_{i\sigma}
c^{\dagger}_{j\sigma^{'}}
c_{l\sigma^{'}}
c_{k\sigma},
\end{aligned}
\end{equation}
where 
$c^{\dagger}_{i \sigma}$ and $c_{i \sigma}$ 
are the creation and annihilation operators for an electron with spin $\sigma$ in a Wannier orbital localized at site $i$,
$T_{ij}$ is the Fourier transform of the band energy
$\epsilon_{\textbf{k}}$
\begin{equation}
\begin{aligned}{}
T_{ij} &=\frac{1}{N} \sum_{\textbf{k}} \epsilon_{\textbf{k}} e^{i\textbf{k}\cdot (\textbf{R}_{i}-\textbf{R}_{j})},
\end{aligned}
\end{equation}
and the Wannier representation matrix element is given by
\begin{equation}\label{eq:wannierelement}
\begin{aligned}{}
\langle ij|\frac{1}{r}|kl\rangle=e^2
\int d \textbf{x} d \textbf{x}^{'} 
\frac{\phi^{*}(\textbf{x}-\textbf{R}_{i})
\phi(\textbf{x}-\textbf{R}_{k})
\phi^{*}(\textbf{x}^{'}-\textbf{R}_{j})
\phi(\textbf{x}^{'}-\textbf{R}_{l})}
{|\textbf{x} - \textbf{x}^{'}|},
\end{aligned}
\end{equation}
where $\phi(\textbf{x}-\textbf{R}_{i})$ and
$\phi^*(\textbf{x}-\textbf{R}_{i})$
are the Wannier functions localized around lattice site $i$. 

In the most general cases where
$i \neq j \neq k \neq l$,
the Wannier matrix elements
$\langle ij|\frac{1}{r}|kl\rangle$
are the so-called four-center integrals, and the corresponding terms in the series of the Coulomb interaction in Eq (\ref{eq:wannier}) can be referred to as the four-site terms.
Since the Wannier function $\phi(\textbf{x}-\textbf{R}_{i})$ goes to zero rapidly when $\textbf{x}$ is away from $\textbf{R}_{i}$, the matrix element (\ref{eq:wannierelement}) is not negligible only when the sites $i,j,k,l$ are close enough so that the overlap between the Wannier functions is sufficiently large.
Therefore, the Coulomb interaction energy can be conveniently approximated by a number of its leading terms of the series in the Wannier representation expression given in Eq (\ref{eq:wannier}).
The biggest interaction term is the so-called on-site term ($i=j=k=l$)
\begin{equation}\label{eq:0a}
\begin{aligned}{}
\frac{1}{2}U \sum_{i,\sigma} n_{i \sigma}n_{i \bar{\sigma}} =
 U \sum_{i} n_{i\uparrow}n_{i\downarrow},
\end{aligned}
\end{equation}
where $U=\langle ii|\frac{1}{r}|ii\rangle$.
If only this on-site term is taken into account, the Hubbard model is obtained.
The next leading order terms are the ones where the set $\{i,j,k,l\}$ contains only two distinct nearest-neighbor sites.
These terms may be referred to as the two-site interaction terms \cite{twosite}.
In this paper, we take one step further beyond the Hubbard model:
In the approximation of the Coulomb interaction, we will retain all of these two-site interaction terms, in addition to the Hubbard on-site term.

\subsection{Two-Site Interaction Approximation}

We now describe a generalized Hubbard model with all of the two-site interaction terms between two nearest-neighbor sites.
The two-side interaction terms are those where the set $\{i,j,k,l\}$ consists of only one pair of nearest-neighbor sites.
These terms fall under two categories:

1. Among the four sites $i,j,k,l$, three of them are in fact the same site, and the fourth one is a nearest neighbor of the other three. There are four such possibilities:

a. $i=j=k$, and $l$ is the n.n. site of $i$:
\begin{equation}\label{eq:1a}
\begin{aligned}{}
\frac{1}{2}\sum_{ij,\sigma} \gamma_{ij}
\langle ii|\frac{1}{r}|ij\rangle
n_{i\sigma}
c^{\dagger}_{i\bar{\sigma}}
c_{j\bar{\sigma}}=
\frac{1}{2}\sum_{ij,\sigma} \gamma_{ij}
\langle ii|\frac{1}{r}|ij\rangle
n_{i\bar{\sigma}}
c^{\dagger}_{i\sigma}
c_{j\sigma}
\end{aligned}
\end{equation}

b. $i=j=l$, and $k$ is the n.n. site of $i$:
\begin{equation}\label{eq:1b}
\begin{aligned}{}
\frac{1}{2}\sum_{ij,\sigma} \gamma_{ij}
\langle ii|\frac{1}{r}|ji\rangle
n_{i\bar{\sigma}}
c^{\dagger}_{i\sigma}
c_{j\sigma}
\end{aligned}
\end{equation}

c. $i=l=k$, and $j$ is the n.n. site of $i$:
\begin{equation}\label{eq:1c}
\begin{aligned}{}
\frac{1}{2}\sum_{ij,\sigma} \gamma_{ij}
\langle ij|\frac{1}{r}|ii\rangle
n_{i\sigma}
c^{\dagger}_{j\bar{\sigma}}
c_{i\bar{\sigma}}=
\frac{1}{2}\sum_{ij,\sigma} \gamma_{ij}
\langle ji|\frac{1}{r}|jj\rangle
n_{j\bar{\sigma}}
c^{\dagger}_{i\sigma}
c_{j\sigma}
\end{aligned}
\end{equation}

d. $j=l=k$, and $i$ is the n.n. site of $j$:
\begin{equation}\label{eq:1d}
\begin{aligned}{}
\frac{1}{2}\sum_{ij,\sigma} \gamma_{ij}
\langle ij|\frac{1}{r}|jj\rangle
n_{j\bar{\sigma}}
c^{\dagger}_{i\sigma}
c_{j\sigma}
\end{aligned}
\end{equation}
Eq (\ref{eq:1a}) is true because the spin summation index $\sigma$ can be replaced by $\bar{\sigma}$ in the expression.
Similarly, the right-hand-side of Eq (\ref{eq:1c}) results from interchanging the summation indexes $i$ and $j$, in addition to the change of $\sigma \rightarrow \bar{\sigma}$.
From Eq (\ref{eq:wannierelement}), it can be verified that all the matrix elements in Eqs (\ref{eq:1a}-\ref{eq:1d}) are equal.
Thus, all of the terms above can be combined into the following expression
\begin{equation}\label{eq:1all}
\begin{aligned}{}
X\sum_{ij,\sigma} \gamma_{ij}
c^{\dagger}_{i\sigma}
c_{j\sigma}
(n_{i\bar{\sigma}}+n_{j\bar{\sigma}}),
\end{aligned}
\end{equation}
where the parameter $X$ represents the matrix elements 
\begin{equation}\label{eq:x}
\begin{aligned}{}
X= \langle ii|\frac{1}{r}|ij\rangle=
\langle ii|\frac{1}{r}|ji\rangle=
\langle ij|\frac{1}{r}|ii\rangle=
\langle ji|\frac{1}{r}|ii\rangle.
\end{aligned}
\end{equation}
Here, $i$ and $j$ represent the nearest-neighbor sites.

2. Among the sites $i,j,k,l$, there are two identical pairs, and moreover, the said two pairs are nearest neighbors.  
There are three such possibilities:

a. $i=j$, $l=k$, $i$ and $l$ are the n.n. sites:
\begin{equation}\label{eq:2a}
\begin{aligned}{}
\frac{1}{2}\sum_{ij,\sigma} \gamma_{ij}
\langle ii|\frac{1}{r}|jj\rangle
c^{\dagger}_{i\sigma}
c_{j\sigma}
c^{\dagger}_{i\bar{\sigma}}
c_{j\bar{\sigma}}
\end{aligned}
\end{equation}

b. $i=k$, $j=l$, $i$ and $j$ are the n.n. sites:
\begin{equation}\label{eq:2b}
\begin{aligned}{}
\frac{1}{2}\sum_{ij,\sigma\sigma'} \gamma_{ij}
\langle ij|\frac{1}{r}|ij\rangle
n_{i\sigma}
n_{j\sigma'}
\end{aligned}
\end{equation}

c. $i=l$, $j=k$, $i$ and $j$ are the n.n. sites:
\begin{equation}\label{eq:2c}
\begin{aligned}{}
\frac{1}{2}\sum_{ij,\sigma\sigma'} \gamma_{ij}
\langle ij|\frac{1}{r}|ji\rangle
c^{\dagger}_{i\sigma}
c^{\dagger}_{j\sigma'}
c_{i\sigma'}
c_{j\sigma}
\end{aligned}
\end{equation}
It has been argued that
$\langle ii|\frac{1}{r}|jj\rangle$ and 
$\langle ij|\frac{1}{r}|ji\rangle$ are in the same order, but much smaller than $\langle ij|\frac{1}{r}|ij\rangle$  \cite{hubbard,strack}.
Therefore, summing over all of the three terms in Eqs (\ref{eq:2a}-\ref{eq:2c}), we have the following terms in this category
\begin{equation}\label{eq:2all}
\begin{aligned}{}
\frac{1}{2} V \sum_{ij,\sigma\sigma'} \gamma_{ij}
n_{i\sigma}
n_{j\sigma'}+
\frac{1}{2} Y \sum_{ij,\sigma} \gamma_{ij} (
c^{\dagger}_{i\sigma}
c_{j\sigma}
c^{\dagger}_{i\bar{\sigma}}
c_{j\bar{\sigma}}
+\sum_{\sigma'}
c^{\dagger}_{i\sigma}
c^{\dagger}_{j\sigma'}
c_{i\sigma'}
c_{j\sigma}
),
\end{aligned}
\end{equation}
where the parameters $V$ and $Y$ represent the matrix elements
\begin{equation}\label{eq:x}
\begin{aligned}{}
V & = \langle ij|\frac{1}{r}|ij\rangle,
\\
Y &=\langle ii|\frac{1}{r}|jj\rangle=
\langle ij|\frac{1}{r}|ji\rangle.
\end{aligned}
\end{equation}
Eqs (\ref{eq:1all}) and (\ref{eq:2all}) contain all the two-site Coulomb interaction terms between two nearest neighbors.

Similarly, for the kinetic energy term, only the constant and nearest-neighbor hopping terms are considered for consistency
\begin{equation}\label{eq:tij}
\begin{aligned}{}
T_{ij}=\frac{1}{N} \sum_{\textbf{k}} \epsilon_{\textbf{k}} e^{i\textbf{k}\cdot (\textbf{R}_{i}-\textbf{R}_{j})}=
T_{0}\delta_{ij}+T_1 \gamma_{ij}+...\, ,
\end{aligned}
\end{equation}
where
\begin{equation}
\begin{aligned}{}
T_0 &=T_{ii}=\frac{1}{N} \sum_{\textbf{k}} \epsilon_{\textbf{k}},
\\
T_1 &=T_{\langle ij \rangle}
= \frac{1}{N} \sum_{\textbf{k}} \epsilon_{\textbf{k}} e^{i\textbf{k}\cdot (\textbf{R}_{i}-\textbf{R}_{j})}, \quad
\mbox{for n.n. sites} \: \langle ij \rangle,
\end{aligned}
\end{equation}
and $(...)$ represents the hopping terms beyond nearest-neighbor sites.
Since $T_1<0$, we will use $t=-T_1$ in the expressions hereafter.
 
Substituting Eq (\ref{eq:tij}) into Eq (\ref{eq:wannier}), and approximating the Coulomb interaction by all the terms given in Eqs (\ref{eq:0a}), (\ref{eq:1all}), and (\ref{eq:2all}), we have
\begin{equation}\label{eq:tpq0}
\begin{aligned}{}
H & \simeq T_{0} \sum_{i,\sigma} n_{i\sigma}
-t\sum_{ij,\sigma}\gamma_{ij}c^{\dagger}_{i \sigma}c_{j \sigma}
+ U \sum_{i} n_{i \uparrow}n_{i\downarrow}
\\
&+\frac{1}{2} V \sum_{ij,\sigma\sigma^{'}} \gamma_{ij} n_{i \sigma}n_{j \sigma^{'}}
+X \sum_{ij,\sigma}\gamma_{ij}c^{\dagger}_{i \sigma} c_{j \sigma}
\big(n_{i \bar{\sigma}}+ n_{j \bar{\sigma}}\big)
\\
&+\frac{1}{2} Y \sum_{ij,\sigma} \gamma_{ij} \big(
c^{\dagger}_{i\sigma}
c_{j\sigma}
c^{\dagger}_{i\bar{\sigma}}
c_{j\bar{\sigma}}
+\sum_{\sigma'}
c^{\dagger}_{i\sigma}
c^{\dagger}_{j\sigma'}
c_{i\sigma'}
c_{j\sigma}\big).
\end{aligned}
\end{equation}
Eq (\ref{eq:tpq0}) contains all the two-site interaction terms as well as the on-site term and can be considered the most complete generalization of the Hubbard model to date.
These two-site terms have been discussed in literature \cite{spalek0,strack,tao}.
The $X$-term is the so-called bond-change interaction, which describes the density-dependent nearest-neighbor hopping of electrons.
The $V$-term is the so-called nearest-neighbor density interaction.
The matrix element $Y$ was discussed in reference \cite{hubbard}.
When $X=V=Y=0$, Eq (\ref{eq:tpq0}) reduced to the Hubbard model.

It is worth pointing out that, in contrast to the $U$-term, the terms parametrized by $V$, $X$, and $Y$ retain the long-range nature of the Coulomb interaction.
Mott argued that, based on the long-range nature of the Coulomb interaction, the correlated electron system undergoes a first-order metal-insulator transition triggered by electronic correlations  \cite{mott}.
The Hubbard model, however, only exhibits a continuous metal-insulator transition owing to the correlation effect of electrons \cite{hubbard}.
The reason that the Hubbard model does not predict a discontinuity in the number of current carriers may be attributed to the lack of the long-range interactions \cite{hubbard,mott,belitz}.
Therefore, a discontinuous metal-insulator transition may exist in the model given in Eq (\ref{eq:tpq0}) \cite{1storderft}.

\subsection{Gutzwiller Projection Technique}

Similar to the Hubbard model, in the model described by Eq (\ref{eq:tpq0}), there are four possible states on the lattice sites: empty, singly occupied with spin up, singly occupied with spin down, and doubly occupied.
The Hamiltonian describing the low-energy physics should not include the doubly occupied states.
We need to project the Hamiltonian (\ref{eq:tpq0}) to a subspace of the Hilbert space where only empty and singly occupied sites are allowed.
This can be accomplished by the Gutzwiller projection approach.
The Gutzwiller projection operators can be defined in the following form \cite{gutzwiller,onevirtualdstate}
\begin{equation}\label{eq:gutz}
\begin{aligned}{}
P_{s} &=\prod_{i}(1-n_{i \uparrow} n_{i \downarrow}),
\\
P_{d} &=I-P_{s},
\end{aligned}
\end{equation}
where
$n_{i \uparrow}$ and $n_{i \downarrow}$ are the electron number operators at site $i$ with spin up and down, respectively, $i$ runs over all the lattice sites, and $I$ is the identity operator.
It can easily be verified that the projection operators $P_s$ and $P_d$ have the following properties
\begin{equation}\label{eq:gutz_p}
\begin{aligned}{}
& (P_{s})^2 =P_{s}, \,\, (P_{s})^n=P_{s},
\\
& (P_{d})^2 =P_{d}, \,\, (P_{d})^n=P_{d},
\\
& P_s P_d =0,
\end{aligned}
\end{equation}
where $n$ is any positive integer.
The Gutzwiller projection operators can partition the original Hilbert space into two subspaces. The first one (denoted by $S$) contains the states with only empty and singly occupied sites, while the second one (denoted by $D$) contains the states with at least one doubly occupied site.
The subscripts $s$ and $d$ denote the two subspaces $S$ and $D$, respectively.

We now derive the effective Hamiltonian in the subspace $S$ using the Gutzwiller projection method.
The specific treatment follows the formalism developed in Ref. \cite{wilson}.
The Schr\"{o}dinger eigenvalue equation of the Hamiltonian $H$ given in Eq (\ref{eq:tpq0}) can be written as
\begin{equation}\label{eq:schr0}
\begin{aligned}{}
H|\Psi\rangle = \varepsilon |\Psi\rangle,
\end{aligned}
\end{equation}
where $|\Psi\rangle$ is the wave vector function in Hilbert space and $\varepsilon$ its corresponding energy eigenvalue.
Since $P_s + P_d = I$, Eq (\ref{eq:schr0}) may be written as
\begin{equation}\label{eq:psd}
\begin{aligned}{}
H (P_s + P_d)|\Psi\rangle = \varepsilon (P_s + P_d) |\Psi\rangle.
\end{aligned}
\end{equation}
Operating $P_s$ and $P_d$ respectively to Eq (\ref{eq:psd}) from the left, and taking the properties of $P_s$ and $P_d$ into account, we obtain
\[
\begin{bmatrix}
P_s H P_s - \varepsilon & P_s H P_d \\
P_d H P_s & P_d H P_d - \varepsilon
\end{bmatrix}
\begin{bmatrix}
|\Psi_s\rangle \\
|\Psi_d\rangle
\end{bmatrix}
=0,
\]
where 
$|\Psi_s\rangle=P_s|\Psi\rangle$ and 
$|\Psi_d\rangle=P_d|\Psi\rangle$, 
the wave vectors in the subspaces $S$ and $D$, respectively.
The projected Hamiltonian matrix above can be diagonalized as 
\[
\begin{bmatrix}
H_s - \varepsilon & 0 \\
0 & H_d - \varepsilon
\end{bmatrix}
\begin{bmatrix}
|\Psi_s\rangle \\
|\Psi_d\rangle
\end{bmatrix}
=0,
\]
where 
\begin{align}\label{eq:hshd}
H_s &=P_s H P_s+P_s H P_d (\varepsilon-P_d H P_d)^{-1}P_d H P_s, 
\\\label{eq:hshd_d}
H_d &=P_d H P_d+P_d H P_s (\varepsilon-P_s H P_s)^{-1}P_s H P_d,
\end{align}
are the projected Hamiltonians in the subspaces. In the following discussion, we are only interested in $H_s$, the effective Hamiltonian in the singly occupied subspace.

Operating the projection operators $P_s$ and $P_d$ on $H$ in Eq (\ref{eq:tpq0}) from both the left and right, after some straightforward manipulations, we can obtain the following projected Hamiltonian matrix elements
\begin{align}
P_s H P_s  =
&T_{0}\sum_{i,\sigma}\tilde{n}_{i\sigma}
-t\sum_{ij,\sigma}\gamma_{ij}
\tilde{c}^{\dagger}_{i \sigma}
\tilde{c}_{j \sigma}
+\frac{1}{2} V \sum_{ij,\sigma\sigma^{'}} \gamma_{ij} 
\tilde{n}_{i \sigma}
\tilde{n}_{j \sigma^{'}} \nonumber
\\
&+\frac{1}{2} Y \sum_{ij,\sigma} \gamma_{ij} \big(
c^{\dagger}_{i\sigma}
c_{j\sigma}
c^{\dagger}_{j\bar{\sigma}}
c_{i\bar{\sigma}}
-\tilde{n}_{i \sigma}
\tilde{n}_{j \sigma}\big), \label{eq:shs}
\\
P_s H P_d =&-(t-X) \sum_{ij,\sigma}\gamma_{ij}
\tilde{c}^{\dagger}_{i \sigma}
c_{j \sigma}
n_{j\bar{\sigma}}, \label{eq:shd}
\\
P_d H P_s =&-(t-X) \sum_{ij,\sigma}\gamma_{ij}
n_{i\bar{\sigma}}
c^{\dagger}_{i \sigma}
\tilde{c}_{j \sigma}, \label{eq:dhs}
\\
P_d H P_d =
&-(t-2X)\sum_{ij,\sigma}\gamma_{ij}
n_{i\bar{\sigma}}
c^{\dagger}_{i \sigma}
c_{j \sigma}
n_{j\bar{\sigma}}
+(U+2T_0) \sum_{i} n_{i \uparrow}n_{i \downarrow} \nonumber
\\
&+(2 V-Y) \sum_{ij}\gamma_{ij}
n_{i\uparrow}
n_{i\downarrow}\big(
n_{j\uparrow}+
n_{j\downarrow}
-
n_{j\uparrow}
n_{j\downarrow}
\big) \nonumber
\\
&+\frac{1}{2} Y \sum_{ij,\sigma}\gamma_{ij}
c^{\dagger}_{i\sigma}
c_{j\sigma}
c^{\dagger}_{i\bar{\sigma}}
c_{j\bar{\sigma}},\label{eq:dhd}
\end{align}
where
$\tilde{c}^{\dagger}_{i \sigma}=
c^{\dagger}_{i \sigma}(1-n_{i\bar{\sigma}})$, 
$\tilde{c}_{i \sigma}=
c_{i \sigma}(1-n_{i\bar{\sigma}})$,
and 
$\tilde{n}_{i}=\sum_{\sigma}
\tilde{c}^{\dagger}_{i \sigma}\tilde{c}_{i \sigma}$
are the projected creation, annihilation, and number operators of electrons.
Note that the $V$- and $Y$- terms do contribute to the zeroth-order of the effective Hamiltonian $P_s H P_s$.
These are the terms neglected in the $t$-$J$ model.
Note also that the $X$ and $U$-terms do not survive in the expression of $P_s H P_s$, indicating that they are not operators of the subspace $S$.
As we will see below, however, they do contribute to the first-order terms in the perturbation expansion when the projected Hamiltonian matrix is diagonalized. 

With the results above, we are now ready to obtain the effective Hamiltonian $H_s$. 
For convenience, we rewrite 
\begin{align}
&H_s = H^{(0)}_s+H^{(1)}_s, \label{eq:hs}
\\
&H^{(0)}_s=P_s H P_s, \label{eq:hs0}
\\
&H^{(1)}_s=P_s H P_d (\varepsilon-P_d H P_d)^{-1}P_d H P_s. \label{eq:hs1}
\end{align}
The zeroth-order term $H^{(0)}_s$ has been obtained in Eq (\ref{eq:shs}).
In order to construct the effective Hamiltonian $H_s$ in the subspace $S$, the key operator to be determined is the inverse operator
$(\varepsilon-P_d H P_d)^{-1}$.
Since this is an operator in the subspace $D$, using the completeness of the wave vectors $|\Psi_{d}\rangle$, we have
\begin{equation}\label{eq:inverse}
\begin{aligned}{}
(\varepsilon-P_d H P_d)^{-1}=\sum_{r,r'}
|\Psi^r_{d}\rangle
\langle\Psi^r_{d}|
(\varepsilon-P_d H P_d)^{-1}
|\Psi^{r'}_{d}\rangle
\langle\Psi^{r'}_{d}|,
\end{aligned}
\end{equation}
where the superscripts $r$ and $r'$ are the labels of the states in the subspace $D$.
Please note that these states in $D$ must be virtual because this inverse operator is in the middle of $H^{(1)}_s$, which is an operator of $S$.
As can be seen from Eq (\ref{eq:dhd}), the expression of $P_d H P_d$ is complicated.
Nevertheless, in the case where there is one virtual doubly occupied site \cite{onevirtualdstate}, the matrix element 
\begin{equation}\label{}
\begin{aligned}{} 
\langle\Psi^r_{d}|
(\varepsilon &-P_d H P_d)^{-1}
|\Psi^{r'}_{d}\rangle =
\langle\Psi^r_{d}|\Big[\varepsilon-2T_0-U'+
\\
&\sum_{ij,\sigma}\gamma_{ij}
c^{\dagger}_{i \sigma}
c_{j \sigma}
\Big(
(t-2X)
n_{i\bar{\sigma}}
n_{j\bar{\sigma}}
-\frac{1}{2} Y
c^{\dagger}_{i\bar{\sigma}}
c_{j\bar{\sigma}}
\Big)
\Big]^{-1}|\Psi^{r'}_{d}\rangle,
\end{aligned}
\end{equation}
where  $U'=U+z' (2V-Y)$, $0 \leq z' \leq z$, and $z$ is the coordinate number.
If we consider the parameters $\varepsilon-2T_0$, $t-2X$, and $Y$ to be small quantities compared with $U'$, we can perform a perturbation expansion with respect to $1/U'$.
To the leading term in $1/U'$, we have
\begin{align}\nonumber
(\varepsilon &-P_d H P_d)^{-1}
\\ \nonumber
&=\sum_{r,r'}
|\Psi^r_{d}\rangle
\langle\Psi^r_{d}|
\Big[-\frac{1}{U'}\Big(1+O\big(\frac{\varepsilon-2T_0}{U'},\frac{t-2X}{U'},\frac{Y}{U'}\big)\Big)^{-1}\Big]
|\Psi^{r'}_{d}\rangle
\langle\Psi^{r'}_{d}|
\\\nonumber
&=
\Big[-\frac{1}{U'}\Big(1+O\big(\frac{\varepsilon-2T_0}{U'},\frac{t-2X}{U'},\frac{Y}{U'}\big)\Big)^{-1}\Big]
\sum_{r,r'}
|\Psi^r_{d}\rangle
\langle\Psi^{r'}_{d}| \delta_{r,r'}
\\\nonumber
&=
-\frac{1}{U'}\Big(1+O\big(\frac{\varepsilon-2T_0}{U'},\frac{t-2X}{U'},\frac{Y}{U'}\big)\Big)^{-1}
\\
&=-\frac{1}{U'}+O\big(\frac{\varepsilon-2T_0}{U^{'2}},\frac{t-2X}{U^{'2}},\frac{Y}{U^{'2}}\big). \label{eq:inverse2}
\end{align}
Substituting (\ref{eq:inverse2}) into Eq (\ref{eq:hs1}) and using the expressions of $P_s H P_d$ and $P_d H P_s$ in Eqs (\ref{eq:shd}) and Eqs (\ref{eq:dhs}), we obtain for the second term of $H_s$ in the form
\begin{equation}\label{eq:j0_0}
\begin{aligned}{}
H^{(1)}_s &= 
-J_0 \sum_{ij,\sigma} \sum_{i'j',\sigma^{'}}
\gamma_{ij}\gamma_{i'j'}
\tilde{c}^{\dagger}_{i \sigma}
c_{j \sigma}
n_{j \bar{\sigma}}
n_{i' \bar{\sigma}'}
c^{\dagger}_{i' \sigma'}
\tilde{c}_{j'\sigma'},
\end{aligned}
\end{equation}
where $J_0=(t-X)^2/U'$.
In general, Eq (\ref{eq:j0_0}) represents a term with four sites $i,j,i',j'$ involved.
In the most general case where $i\neq j\neq i'\neq j'$, it is a four-site term and an operator of the subspace $D$.
Therefore we examine its special cases where two of the four sites are in fact the same site.
Since $<ij>$ and $<i'j'>$ are two nearest-neighbor pairs, there are only four such possibilities:\\

1. $i$ and $i'$ are the same site

2. $i$ and $j'$ are the same site

3. $j$ and $i'$ are the same site

4. $j$ and $j'$ are the same site\\

\noindent
Further examination reveals that cases $1$ and $4$ result in a zero operator, while case $2$ is still an operator in the subspace $D$.
Only case 3 with $i'=j$ provides a meaningful contribution to the operator in the subspace $S$
\begin{equation}\label{eq:j0_1}
\begin{aligned}{}
H^{(1)}_{s}
&=-J_0 \sum_{ijk,\sigma} \gamma_{ij}\gamma_{jk}
\tilde{c}^{\dagger}_{i \sigma} \big(
\tilde{n}_{j \bar{\sigma}}
\tilde{c}_{k \sigma}
+
c_{j \sigma}
c^{\dagger}_{j \bar{\sigma}}
\tilde{c}_{k \bar{\sigma}}\big).
\end{aligned}
\end{equation}
Please note that in Eq (\ref{eq:j0_1}) the site $j$ is the nearest neighbor of both sites $i$ and $k$.
Therefore, the two-site term can only result from the case of $i=k$ in the summation.
Writing the two-site and three-site terms separately, we have
\begin{equation}\label{eq:j0_1b}
\begin{aligned}{}
&H^{(1)}_{s}=(H^{(1)}_{s})_{2site}+(H^{(1)}_{s})_{3site},
\end{aligned}
\end{equation}
where
\begin{align}{}
&(H^{(1)}_{s})_{2site} =-J_0 \sum_{ij,\sigma} \gamma_{ij}\big(
\tilde{n}_{i \sigma}
\tilde{n}_{j \bar{\sigma}}
+
c^{\dagger}_{i \sigma}
c_{j \sigma}
c^{\dagger}_{j \bar{\sigma}}
c_{i \bar{\sigma}}\big), \label{eq:hs_2}
\\
&(H^{(1)}_{s})_{3site} =-J_0 \sum_{i\neq k,j,\sigma} \gamma_{ij}\gamma_{jk}
\tilde{c}^{\dagger}_{i \sigma}\big(
\tilde{n}_{j \bar{\sigma}}
\tilde{c}_{k \sigma}
+
c_{j \sigma}
c^{\dagger}_{j \bar{\sigma}}
\tilde{c}_{k \bar{\sigma}}\big). \label{eq:hs_3}
\end{align}
Please note that the summation in Eq (\ref{eq:hs_3}) excludes the case $i=k$.
This three-site term is usually ignored in the discussion of the $t$-$J$ model.
We will also neglect this three-site term in the subsequent discussions.
 
Substituting the expression of $H^{(0)}_s$ in Eq (\ref{eq:shs}) and
$H^{(1)}_s \simeq (H^{(1)}_{s})_{2site} $ in Eq (\ref{eq:hs_2}) into Eq (\ref{eq:hs}), we finally obtain, to the order of two-site interactions terms, the effective Hamiltonian $H_s$
\begin{equation}\label{eq:tpq10}
\begin{aligned}{}
H_s &=
T_{0} \sum_{i\sigma} \tilde{n}_{i\sigma}+
\sum_{ij,\sigma}\gamma_{ij} \Big(-t \, \tilde{c}^{\dagger}_{i \sigma}\tilde{c}_{j \sigma}
- \frac{1}{2} (2J_0-Y)\, 
c^{\dagger}_{i\sigma}
c^{\dagger}_{j \bar{\sigma}}
c_{i \bar{\sigma}}
c_{j\sigma}
\\
&+ \frac{1}{2} (V-Y)\, \tilde{n}_{i \sigma} \tilde{n}_{j \sigma}
+ \frac{1}{2} (V-2J_0)\, \tilde{n}_{i \sigma}\tilde{n}_{j \bar{\sigma}}\Big),
\end{aligned}
\end{equation}
which can be written as
\begin{equation}\label{eq:tpq1}
\begin{aligned}{}
H_s &=
T_{0} \sum_{i\sigma} \tilde{n}_{i\sigma}+
\sum_{ij,\sigma}\gamma_{ij} \Big(-t \, \tilde{c}^{\dagger}_{i \sigma}\tilde{c}_{j \sigma}
- \frac{1}{2} J\, 
c^{\dagger}_{i\sigma}
c^{\dagger}_{j \bar{\sigma}}
c_{i \bar{\sigma}}
c_{j\sigma}
\\
&+ \frac{1}{2} p\, \tilde{n}_{i \sigma} \tilde{n}_{j \sigma}
+ \frac{1}{2} q\, \tilde{n}_{i \sigma}\tilde{n}_{j \bar{\sigma}}\Big),
\end{aligned}
\end{equation}
with the coefficients given by the following equations
\begin{equation}\label{eq:pars}
\begin{aligned}{}
p &=V-Y,
\\
q &=V-2 J_{0}, \,\,
J_{0} =(t-X)^2/U',\,\, U'=U+z'(2V-Y),
\\
J &=p-q,
\end{aligned}
\end{equation}
where $0 \leq z'\leq z$ as discussed above.
Eq (\ref{eq:tpq1}) is the effective Hamiltonian in the subspace of singly occupied states.
Since it is projected from a more general Hamiltonian retaining all terms up to the two-site interactions, we believe that Eq (\ref{eq:tpq1}) can describe the low-energy physics of the interacting electron systems in solids more accurately and completely.
Moreover, there is no precondition imposed on the electron density, $n$, so the above derivation is applicable to systems at any doping level.

Please note that by rescaling $H_s$ with $t$, this model has only three dimensionless independent parameters $\bar{T}_0=T_0/t$, $\bar{p}=p/t$, and $\bar{q}=q/t$. 
One may expect that the system will exhibit ferromagnetism when $\bar{q}$ is sufficiently larger than $\bar{p}$. 
On the other hand, the system will exhibit antiferromagnetism when $\bar{p}$ is sufficiently larger than $\bar{q}$.
Our studies demonstrate that this is true.
More interestingly, our study also shows that this model exhibits superconductivity for some parameter range of $\bar{p} > \bar{q}$.

\section{The $t$-$J$ Model}

Apparently, Eq (\ref{eq:tpq1}) reduces to the $t$-$J$ model with $J=2J_0$ in the case of $X=Y=V=0$.
We will show that when $Y=V$, Eq (\ref{eq:tpq1}) will reduce to the $t$-$J$ model as well (with $J=2J_0-V$ in this case). 
Making use of the following operator identities
\begin{align}{} \label{eq:rel0}
&C^{\dagger}_{i\uparrow}C_{i\downarrow}=S^{+}_{i}
=S^{x}_{i}+iS^{y}_{i},
\\
&C^{\dagger}_{i\downarrow}C_{i\uparrow}=S^{-}_{i}
=S^{x}_{i}-iS^{y}_{i},
\\
&n_{i\uparrow}-n_{i\downarrow}=2S_{i}^{z},
\end{align}
we obtain
\begin{align}{} \label{eq:rel1}
&\sum_{\sigma}
c^{\dagger}_{i \sigma}
c_{j \sigma}
c^{\dagger}_{j \bar{\sigma}}
c_{i \bar{\sigma}}
=-2\big(S_i^x S_j^x+S_i^y S_j^y\big),
\\ \label{eq:rel2}
&\sum_{\sigma}\big(
\tilde{n}_{i\sigma}
\tilde{n}_{j\sigma}-
\tilde{n}_{i\sigma}
\tilde{n}_{j\bar{\sigma}}\big)=4 S_i^z S_j^z,
\end{align}
where 
$S_i^\alpha$, $\alpha=x,y,z$, are the spin operators at site $i$.
Furthermore, it is easy to verify that
\begin{align}{}
\label{eq:rel3}
&\sum_{\sigma}\big(
\tilde{n}_{i\sigma}
\tilde{n}_{j\sigma}+
\tilde{n}_{i\sigma}
\tilde{n}_{j\bar{\sigma}}\big)=
\tilde{n}_{i}
\tilde{n}_{j},
\end{align}
where 
$\tilde{n}_{i}=\sum_{\sigma}\tilde{n}_{i\sigma}$.
Eqs (\ref{eq:rel1})-(\ref{eq:rel3}) immediately lead to the following well-known relation
\begin{align}{}
\sum_{\sigma}
\big(
c^{\dagger}_{i \sigma}
c_{j \sigma}
c^{\dagger}_{j \bar{\sigma}}
c_{i \bar{\sigma}}
+
\tilde{n}_{i \sigma}
\tilde{n}_{j \bar{\sigma}}
\big)
=-2(S_{i}\cdot S_{j}
-\frac{1}{4}
\tilde{n}_{i}
\tilde{n}_{j}).\label{eq:charge_spin}
\end{align}
Writing the two terms in the summation over $\sigma'$ explicitly, the zeroth-order effective Hamiltonian Eq (\ref{eq:shs}) can be written as
\begin{align}\nonumber
H_s^{(0)}&=P_s H P_s 
=T_{0}\sum_{i,\sigma}\tilde{n}_{i\sigma}
\\\nonumber
&+\sum_{ij\sigma}\gamma_{ij}\Big(
-t\tilde{c}^{\dagger}_{i \sigma}\tilde{c}_{j \sigma}
+\frac{1}{2} Y \big(
c^{\dagger}_{i\sigma}
c_{j\sigma}
c^{\dagger}_{j\bar{\sigma}}
c_{i\bar{\sigma}}
+
\tilde{n}_{i \sigma}
\tilde{n}_{j \bar{\sigma}}\big)
\\
&
+\frac{1}{2} (V-Y) \big(
\tilde{n}_{i \sigma}
\tilde{n}_{j \sigma}+
\tilde{n}_{i \sigma}
\tilde{n}_{j \bar{\sigma}}\big)
\Big). \label{eq:shs_2}
\end{align}
Therefore, using Eqs (\ref{eq:rel3}) and (\ref{eq:charge_spin}), Eq (\ref{eq:shs_2}) takes the form
\begin{align}\nonumber
H_s^{(0)}&=
T_{0}\sum_{i,\sigma}\tilde{n}_{i\sigma}
-t\sum_{ij,\sigma}\gamma_{ij}
\tilde{c}^{\dagger}_{i \sigma}\tilde{c}_{j \sigma}
+\sum_{ij}\gamma_{ij}\Big(
-Y \big(
S_{i}\cdot S_{j}
-\frac{1}{4}
\tilde{n}_{i}
\tilde{n}_{j}\big)
\\
&+
\frac{1}{2} (V-Y)\,\,
\tilde{n}_{i}
\tilde{n}_{j}
\Big).\label{eq:shs_3}
\end{align}
We point out again that the terms parametrized by  $Y$ and $(V-Y)$ in Eq (\ref{eq:shs_3}) are the neglected terms in the $t$-$J$ model.
Similarly, for the first-order term $(H^{(1)}_{s})_{2site}$ given in Eq (\ref{eq:hs_2}), we have 
\begin{equation}
\begin{aligned}{}
(H^{(1)}_{s})_{2site}
&=2J_0 \sum_{ij} \gamma_{ij}(
S_{i}\cdot S_{j}
-\frac{1}{4}
\tilde{n}_{i}
\tilde{n}_{j}
).\label{eq:hs1_2}
\end{aligned}
\end{equation}
Summing up the terms in Eqs (\ref{eq:shs_3}) and (\ref{eq:hs1_2}), the effective Hamiltonian can be expressed as in the $t$-$J$ model form 
\begin{equation}\label{eq:tpq2}
\begin{aligned}{}
H_s &=
T_{0} \sum_{i\sigma} \tilde{n}_{i\sigma}
-t \,\sum_{ij,\sigma}\gamma_{ij} \tilde{c}^{\dagger}_{i \sigma}\tilde{c}_{j \sigma}
+\sum_{ij}\gamma_{ij}\Big(
(2J_0-Y)S_{i}\cdot S_{j}
\\
&-\frac{1}{4}(2J_0+Y-2V)
\tilde{n}_{i}
\tilde{n}_{j}
\Big).
\end{aligned}
\end{equation}
The above equation shows that when $V=Y$, the model reduces to the $t$-$J$ model with $J=2J_0-V$. 
However, it has been argued that $V \gg Y$ in the usual cases \cite{hubbard}.
Also, since the value of $V$ is usually large, the condition of $J=2J_0-V>0$ (necessary for antiferromagnetic and superconducting states) is difficult to be realized.
Therefore, the condition $V=Y$ may be rarely physically satisfied. 
Finally, using the parameters $p$ and $q$ defined in Eq (\ref{eq:pars}), the model can be written as 
\begin{equation}\label{eq:tpq_3}
\begin{aligned}{}
H_s &=T_{0} \sum_{i\sigma} \tilde{n}_{i\sigma}
-t \,\sum_{ij,\sigma}\gamma_{ij} \tilde{c}^{\dagger}_{i \sigma}\tilde{c}_{j \sigma}
+\sum_{ij}\gamma_{ij}\Big[
J\, S_{i}\cdot S_{j}
+\frac{1}{4}J'\,
\tilde{n}_{i}
\tilde{n}_{j}
\Big],
\end{aligned}
\end{equation}
where $J=p-q$ and $J'=p+q$.
When $J'=-J$, which means $p=0$ (i.e. the condition $V=Y$), one has the $t$-$J$ model.
A comparison of Eqs (\ref{eq:tj}) and (\ref{eq:tpq_3}) reveals how the inclusion of the two-site nearest-neighbor Coulomb interaction terms changes the $t$-$J$ model:
The coefficient of the term $\tilde{n}_{i}\tilde{n}_{j}$ changes from $-J/4$ to a positive independent coefficient $J'=p+q$.
This is the only change, and no new terms are generated.
Our Green function analysis shows that a reasonable superconductivity dome appears at a finite value of $p$ ($p/q \sim 2.5$), which again seems to indicate that the coefficient of $\tilde{n}_{i}\tilde{n}_{j}$ 
of the conventional $t$-$J$ model, $-J/4$, is not in the physical range.

Eqs (\ref{eq:tpq1}) and (\ref{eq:tpq_3}) are two alternative expressions of the new model, which reveal the origin of the spin correlations from two different though equivalent perspectives.
For two nonempty nearest-neighbor sites, there are two different spin configurations: the spin directions are either the same or the opposite.
Eq (\ref{eq:tpq1}) indicates that the energies for the two spin configurations are, in general, different ($p\neq q$), and that the relative direction of the spins on the nearest-neighbor sites is determined by the competition between the energies of the two spin configurations.
On the other hand, as shown in Eq (\ref{eq:tpq_3}), if the interaction terms are rearranged such that the amount of energy 
$J S_{i}^{z} S_{j}^{z}$ 
from the $p$- and $q$-terms is combined into the $J$-term to form a complete Heisenberg interaction term \cite{direct}, the energies of the two spin configurations become the same, namely, $J'\tilde{n}_{i}\tilde{n}_{j}/4$.
Then the spin correlations can be attributed to the Heisenberg interaction.

The magnetic ordering of the model is determined by the coefficient of the Heisenberg term, $J=p-q$.
A natural question arises: What if $p \simeq q$?
When $p=q$, Eq (\ref{eq:tpq1}) or (\ref{eq:tpq_3}) becomes
\begin{equation}\label{eq:tpq_00}
\begin{aligned}{}
H &=T_{0} \sum_{i\sigma} \tilde{n}_{i\sigma}
-t \,\sum_{ij,\sigma}\gamma_{ij} \tilde{c}^{\dagger}_{i \sigma}\tilde{c}_{j \sigma}
+\frac{1}{2}p\,\sum_{ij}\gamma_{ij}
\tilde{n}_{i}
\tilde{n}_{j}.
\end{aligned}
\end{equation}
As the results suggest, there should be neither long-range nor short-range magnetic orders in the model described in Eq (\ref{eq:tpq_00}).
This simplified model describes solely the electrical properties of the interacting electron systems. 
Please note that Eq (\ref{eq:tpq_00}) is a Hamiltonian in the subspace of singly occupied states, which is different from the usual generalized Hubbard model with the $V$-term.

\section{Conclusion}

Starting with a generalized Hubbard model with the next leading order Coulomb interaction terms, we have systematically developed a 
new model for the strongly correlated electrons in solids using the Gutzwiller projection approach.
Our study demonstrates that the origin of the Heisenberg interaction term can be attributed to the competition between the energies of the two spin configurations of the nearest-neighbor electrons.
A comparison of the conventional $t$-$J$ model and the new model has been discussed.
It seems that the coefficient of one of the interaction terms in the $t$-$J$ model, i.e. $\tilde{n}_{i}\tilde{n}_{j}$, is not in the physical range, which may be the reason why the superconductivity dome has not been obtained from the $t$-$J$ model.

We have studied the magnetism and superconductivity of this new model. 
The main results of this research have been summarized in a recent paper \cite{tao2}.
Our study indicates that this model provides a more complete description of the physics of strongly correlated electron systems.
The system is not necessarily in a ferromagnetic state as temperature $T\rightarrow 0$ at any doping level $\delta\geq 0$.
The system, however, must be in an antiferromagnetic state at the origin of the doping-temperature ($\delta$-$T$) plane ($T\rightarrow 0$, $\delta=0$).
Moreover, the model exhibits superconductivity in a doped region at sufficiently low temperatures.
The details of the magnetic and superconducting properties of this model will be presented in forthcoming papers.

Finally, since Eq (\ref{eq:tpq_3}) and the $t$-$J$ model differ only by the coefficient of the term $\tilde{n}_{i}\tilde{n}_{j}$, most of the existing numerical tools (computer programs) developed for the $t$-$J$ model can be directly applied to Eq (\ref{eq:tpq_3}) without any major changes.


\begin{thebibliography}{9}
\bibitem{fock}
The Hilbert space describing an electron system where the number of electrons changes is sometimes called the fermionic Fock space. We still prefer to use the more general term, the Hilbert space, in this paper.
\bibitem{1storderft}
In Ref \cite{tao}, we have indicated that the $X$-term alone can trigger the first-order metal-insulator phase transition.
\bibitem{direct}
A direct way to go from Eq (\ref{eq:tpq1}) to Eq (\ref{eq:tpq_3}) is by rearranging the interaction terms with the following: 1) subtract $J S_{i}^{z} S_{j}^{z}$ from the $p$- and $q$-terms, and 2) add it to the $J$-term.
\end{thebibliography}
\end{document}